\documentclass[aps,twocolumn,showkeys,superscriptaddress,floatfix]{revtex4}
\usepackage{upgreek}
\usepackage{graphicx}       
\usepackage{color}
\usepackage[normalem]{ulem} 
\usepackage{amsmath}
\usepackage{amssymb}

\newcommand{\st}{$\rm ^\circ$}     
\newcommand{\stc}{$\rm ^\circ C$}  

\newcommand{\muvk}{$\mu$V/K}

\newcommand{\emucm}{emu/cm$^{3}$}

\newcommand{\eps}{$\varepsilon$-Fe$_2$O$_3$}
\newcommand{\epsx}{$\varepsilon$-Fe$_{1-x}$Al$_x$O$_3$}
\newcommand{\epsal}{$\varepsilon$-Fe$_{1.7}$Al$_{0.3}$O$_3$}
\newcommand{\feomagh}{$\gamma$-Fe$_2$O$_3$}
\newcommand{\feomagn}{Fe$_3$O$_4$}

\begin{document}
\sloppy
\title{Spin Seebeck effect in $\varepsilon$-Fe$_2$O$_3$ thin layer with high coercive field}
\author{K. Kn\'{\i}\v{z}ek}
\author{M. Pashchenko}
\author{P. Levinsk\'{y}}
\author{O. Kaman}
\author{P. Ji\v{r}\'{\i}\v{c}ek}
\author{J. Hejtm\'{a}nek}
\affiliation{
 Institute of Physics of the CAS, Cukrovarnick\'{a} 10, 162 00 Prague 6, Czech Republic.}
\author{M. Soroka}
\author{J. Bur\v{s}\'{\i}k}
\affiliation{
 Institute of Inorganic Chemistry of the CAS, 250 68 \v{R}e\v{z}, Czech Republic.}
\email[corresponding author: ]{knizek@fzu.cz}
\begin{abstract}
Spin Seebeck effect has been investigated in Pt/$\varepsilon$-Fe$_2$O$_3$ bilayers. The $\varepsilon$-Fe$_2$O$_3$ thin layer with $40-70$~nm thickness were deposited by a spin-coating method on Y:ZrO$_2$(100) substrates. The prepared layers are highly oriented with the easy magnetic $a$-axis parallel to the film surface. The magnetic hysteresis loops measured at room temperature with magnetic field parallel to the layer exhibit coercive fields up to 11.6~kOe, which is so far the highest value measured for $\varepsilon$-Fe$_2$O$_3$ thin layer samples. The shape of the spin Seebeck hysteresis loops is similar to the shape of magnetization for single phase layers with coercive field around 10~kOe. In some prepared layers a small amount of secondary soft ferrimagnetic phase is revealed by a constricted shape of magnetization loops, in contrast to spin Seebeck loops, where no constriction is observed.
A difference in encountered in the case of layers with a small amount ($1-2$~volume\%) of secondary soft ferrimagnetic phase, which is revealed by a constricted shape of magnetization loops, in contrast to spin Seebeck loops, where no constriction is observed.
\end{abstract}
%
%
\maketitle
\section{Introduction}


The recently emerged field of spin caloritronics, a multidisciplinary field combining thermoelectricity and spintronic, is concerned with the interplay between heat currents and spin angular momentum transport \cite{RefBoona2014EESCI7_885}.
One of the well-established phenomena of spin caloritronics is the spin Seebeck effect (SSE) discovered in 2008 by Uchida \textit{et al.} \cite{RefUchida2008NAT455_778}. The SSE is a two-step process that requires an interface between a spin-polarized material, typically ferro- or ferrimagnet (FM), and a normal metal (NM). Temperature gradient applied to the FM material generates spin currents, which is converted to electrical current by means of the inverse spin Hall effect (ISHE) \cite{RefSaitoh2006APL88_182509} in the attached NM thin layer. A necessary condition for the generation of SSE is that the directions of the spin current, magnetic moments and electrical current have mutually perpendicular components. In analogy with Seebeck effect, the spin Seebeck coefficient is defined as $S_{SSE} = E_{ISHE} / \nabla T$.


In the this work we present a study of SSE in \eps.
The \eps\ phase was first observed by Forestier and Guiot-Guillain in 1934 \cite{RefForestier1934CRAS199_720} and named by Schrader and Buttner in 1963 \cite{RefSchrader1963ZAACH320_220}.
In distinction to well-known iron oxides, which crystallize in a high symmetric space groups, namely either cubic $Fd\overline{3}m$ in the case of magnetite \feomagn\ and maghemite \feomagh, or hexagonal $R\overline{3}c$ in the case of hematite $\alpha$-Fe$_2$O$_3$, the symmetry of \eps\ is orthorhombic.
The iron atoms occupy in the \eps\ structure four crystallographically inequivalent cation sites: three octahedral and one tetrahedral site, determining the magnetic behaviour of the material. Each of the four inequivalent magnetic sublattices exhibits different temperature dependence \cite{RefOhkoshi2009JPCC113_11235}. The material is ferrimagnetic at room temperature, with relatively low remanence $\sim$8~\emucm\ and saturation magnetization $\sim$17~\emucm, and giant coercive field $\sim$20~kOe ascribed to its high magnetocrystalline anisotropy \cite{RefJin2004ADVMAT16_48,RefTseng2009PRB79_94404}, which can enhanced up to 31~kOe by Rh doping \cite{RefNamai2012NATCOM3_1035}. It undergoes a transition to the paramagnetic state at $T_C\sim$490~K, and a two step magnetic transition in the temperature range $100-150$~K accompanied by a decrease in coercive field \cite{RefPopovici2004CHEMMAT16_5542,RefTseng2009PRB79_94404,RefKohout2015JAP117_17D505}.
The magnetization can be enhanced by a substitution of non-magnetic cation, \textit{e.g.} Al or Ga, at the tetrahedral site, however at the expense of lower coercive field \cite{RefCorbellini2017SCRMAT140_63}.

The high coercive field and multiferroic behaviour suggest the enormous application potential of \eps\ as a functional magnetic material, especially for applications where materials with considerable magnetic hardness is needed \textit{e.g.} in storage media, or devices controlled by external magnetic or electric field.
Coupling of the magnetic and dielectric properties of \eps\ was reported in \cite{RefGich2006NANOTECH17_687}, demonstrating also its multiferroic character.
Another application as an electromagnetic wave absorber for high-speed wireless communication has been proposed in  \cite{RefNamai2009JACHS131_1170}. The frequencies of the absorption peaks for \epsx, originated due to the natural resonance achieved by the large magnetic anisotropies, were observed between $182-112$ GHz for $x=0-0.4$, respectively. Such frequencies are the highest ones for magnetic materials.

For its low surface energy, the dark brown polymorph \eps\ only exists in nanocrystalline forms.
Due to its thermodynamic metastability, it is difficult to prepare \eps\ in a pure form without admixtures of other iron(III) oxide polymorphs. A widely respected synthesis method to achieve high phase purity employs the mesoporous silica template, which prevents the aggregation of particles and provides the space confinement during their growth enhancing thus their thermal stability \cite{RefBrazda2014CGD14_1039}.
So far, \eps\ has been prepared in three morphological forms:
nanoparticles (in sizes $10-200$~nm),
nanorods and nanowires with typical width $10-120$~nm and length more than $200-800$~nm \cite{RefJin2004ADVMAT16_48,RefSakurai2008JPCC112_20212,RefOhkoshi2016SCIREP6_27212}, and
epitaxial films were prepared by Pulsed layer (PLD) with thickness limited to $\sim$110~nm \cite{RefGich2010APL96_112508,RefGich2014ADVMAT26_4645,RefThai2016JAP120_185304,RefCorbellini2017SCIREP7_3712,RefCorbellini2017SCRMAT140_63}. on Al$_2$O$_3$(001), SrTiO$_3$(111) and Y:ZrO$_2$(100) substrates. The thin layers exhibit lower coercive field then reported  for samples in the form of nanoparticles, with so far maximum value 8~kOe.
Thin films prepared by Atomic layer deposition (ALD) with higher thickness 260~nm was reported in \cite{RefTanskanen2017APLMAT5_56104}, however they had a very low coercive field 1.6~kOe.


Due to the large magnetocrystalline anisotropy the induced magnetization has a preferred orientation within the crystal structure with an easy axis of magnetization in the $a$-direction. Owing to their direction of easy growth lying in the $ab$-plane, \eps\ inherently tends to crystallize with their $c$-axis perpendicular to the substrate surface when deposited as thin films. Since the magnetization vector in the SSE element should lie parallel to the film surface, \eps\ is naturally suitable for the SSE experiment.

In this work we present SSE and magnetic measurement of \eps\ and Al-substituted \epsal\ thin films with thickness of 40 and 70~nm deposited on single crystal ($h00$)-oriented yttrium-stabilized zirconia (YSZ) substrates.

\section{Experimental}

\begin{figure}
\centering
\includegraphics[width=1.00\columnwidth,viewport=0  360 535 780,clip]{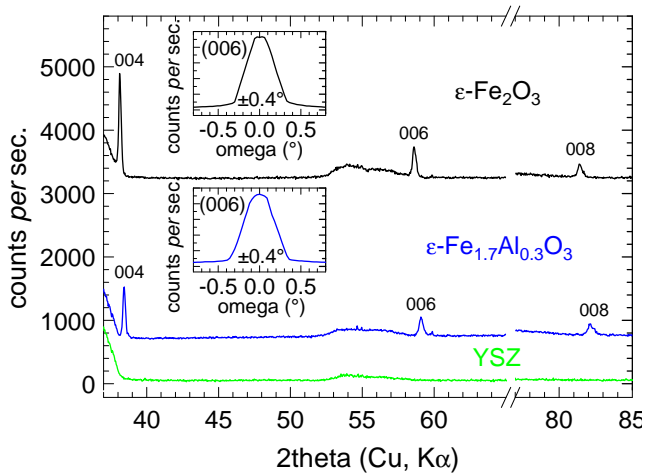} 
\caption{X-ray diffraction of the \eps\ (black line) and \epsal\ (blue line) 70~nm thin films. The insets show rocking-curve measurements. The diffraction peak (400) of the YSZ substrate is skipped.} \label{Fig_XRD}
\end{figure}

\begin{figure}
\centering
\includegraphics[width=1.00\columnwidth,viewport=0  380 535 780,clip]{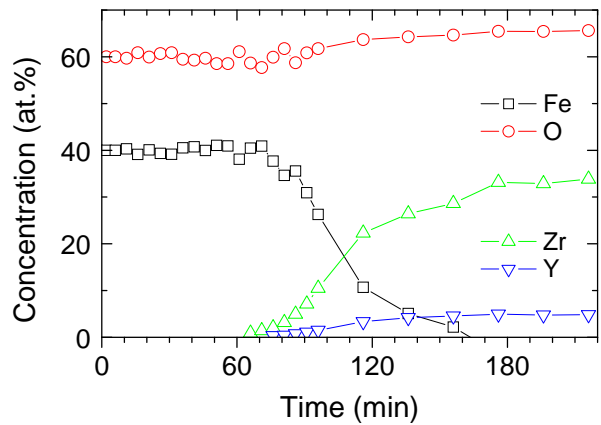} 
\caption{Depth profile of the atom concentrations determined by ESCA for \eps-70~nm.} \label{Fig_ESCA}
\end{figure}


Thin films of iron oxide \epsx\ with $x=0$ and 0.3 were prepared by chemical solution deposition (CSD) method on ($h00$)-oriented, epitaxially polished cubic yttrium-stabilized zirconia (YSZ) with 9.5~mol\% Y$_2$O$_3$ (Crystal GmbH, Germany). Home synthesized iron (III) isobutoxide and aluminum isobutoxide purchased from Sigma Aldrich were used as metal precursor. Deposition solution were formed by dissolving of appropriate amounts of metal alkoxides in iso-butanol. The overall metal concentration in deposition solutions was c(Fe$_{2+}$+Al$_{3+}$)~$=0.076$~mol$\cdot$dm$^{-3}$. In order to prevent reaction with air humidity all reactions and handling were done under dry nitrogen. Before fabrication process, the substrates were cleaned using isopropanol for eliminating any dust or contaminants, and then rinsed with deionized water. Prior to the deposition, substrates were treated with plasma (Zepto Plasma cleaner, Diener Electronic, Germany).

Thin film was deposited by spin-coating technique (spin coater RC8 Gyrset by KarlSuss). Deposition solution was dropped onto the substrate and spun at 3000~rpm for 30s. Thin film was dried at 110\stc\ for several minutes and pyrolyzed at 250\stc\ for 5~min. Crystallization annealing was done at 800\stc\ for 15~min in a conventional tube furnace under open air atmosphere. Deposition-crystallization step was repeated 14~times, until required film thickness 70~nm were reached. Final annealing at 800\stc\ for 60~min was done under the same conditions.

Simultaneously with samples on YSZ, thin films on Si wafers were prepared under the same deposition conditions. These samples served for evaluation of film thicknesses by means of stylus apparatus (AlphaStep IQ, KLA Tencor). The chemical composition of \epsx\ deposition solutions were evaluated by XRF analysis (Rigaku NEX CG energy dispersive fluorescence spectrometer with Rh 50 W X-ray source) on separate powder samples prepared by pyrolysis and annealing of deposition solution.




SSE was measured using home-made apparatus. A longitudinal configuration was used, in which the directions of the spin current, magnetic moments and electrical current are mutually perpendicular, see the schema displayed in Fig.~\ref{Fig_SSE}d. AlN plate, as a thermally conducting and electrically insulating material, was used to separate the heater and the sample in order to uniformly spread the heat flow over the sample area.
The width of the measured sample was 2~mm, the length 7~mm and electric contact distance (on Pt-layer) was 5~mm. Pt layer converting spin current to voltage was deposited using K550X Quorum Technologies sputter coater. The electric resistance of the Pt-layer measured by a 2-point technique varied for respective samples in the series within the range $100-200$~$\Omega$. By comparison with the resistivity of Pt films with variable thickness determined in Ref.~\cite{RefAgustsson2008APSS254_7356}, we estimate the effective thickness of our Pt films as $6-9$~nm.
The resistance of the \eps\ thin layer itself was more than 1~G$\Omega$.
Therefore, the contribution of Nernst effects to the measured signal can be considered as negligible due to the lack of free charge carriers in the \eps.


The phase purity and degree of preferred orientation of the thin films was checked by X-ray diffraction over the angular range $2\uptheta$ from 10 to 120\st\ using the X-ray powder diffractometer Bruker D8 Advance (CuK$\alpha_{1,2}$ radiation, secondary graphite monochromator).
The magnetic response of the samples was measured within the range of magnetic field from $-50$ to 50~kOe at room temperature using a SQUID magnetometer (MPMSXL, Quantum Design)


\section{Results and discussion}


The X-ray diffraction of the thin films only revealed the reflections of the YSZ substrate and ($00l$) reflections of the \epsx\ phases. The $c$-axis preferred orientation is quantified by the full-width at the half-maximum (FWHM) of the rocking curve, which is approx. 0.4\st\ for all thin films, see \ref{Fig_XRD}.
The $c$-lattice parameters of \eps\ samples, 9.462(1)~\AA\ (70~nm) and 9.456(1)~\AA\ (40~nm), calculated using $\cos\uptheta/\tan\uptheta$ extrapolation to correct a possible off-centre position of the film during XRD measurement, are in good agreement with literature values for \eps\ \cite{RefSakurai2005JPSJ74_1946,RefKohout2015JAP117_17D505}.
The stoichiometry of Al-substituted layer was determined as $x=0.30(2)$ by XRF. This in accordance with the $c$-parameter 9.391(1)~\AA\ calculated for \epsx, which agrees well with interpolated value for $x=0.3$ between \eps\ and AlFeO$_3$ \cite{RefBouree1996ACB52_217,RefCorbellini2017SCRMAT140_63}.


The depth profile of atom concentrations determined by ESCA in dependence on the time of ion beam etching for \eps-70~nm thin layer is displayed in \ref{Fig_ESCA}. The analysis revealed a constant content of Fe and O for the first range of the depth profile, corresponding to the Fe$_2$O$_3$ stoichiometry. It is followed by a gradual decrease of Fe content and by a simultaneous increase of Zr and Y accompanied by an increase of oxygen content corresponding to oxygen stoichiometry of Y:ZrO$_2$. This gradual change of cationic content might have been indication of a partial substitution of Zr and Y in \eps\ at the interface between film and substrate. However, inspection of the etched region of the film revealed, that the rate of etching was not uniform throughout the etched area. Therefore the gradual change of cationic content is caused by a simultaneous measurement of the film and the substrate at various parts of the etched area, and not by the substitution.
The ratio of Y:Zr is strictly constant for all measured levels, and as such double substitution in \eps\ is less probable, it also excludes the possible substitution of substrate cations to \eps.


\begin{figure}
\centering
\includegraphics[width=1.00\columnwidth,viewport=0   0 535 780,clip]{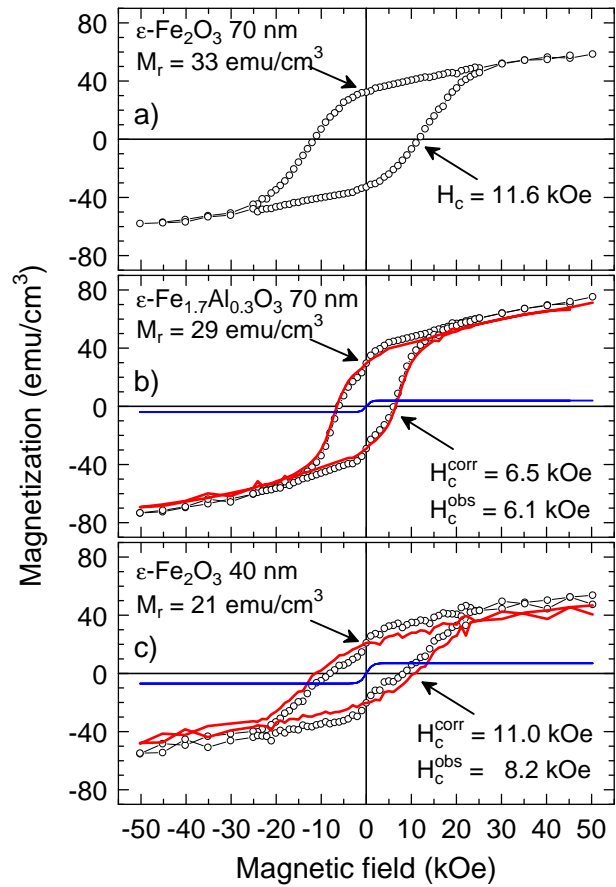} 
\caption{Magnetization hysteresis loops measured at room temperature of a) \eps-70~nm, b) \epsal-70~nm and c) \eps-40~nm. Correction for secondary soft ferrimagnetic phase is displayed.} \label{Fig_Mag}
\end{figure}

\begin{figure*}
\centering
\includegraphics[width=1.00\textwidth,viewport=0 440 565 780,clip]{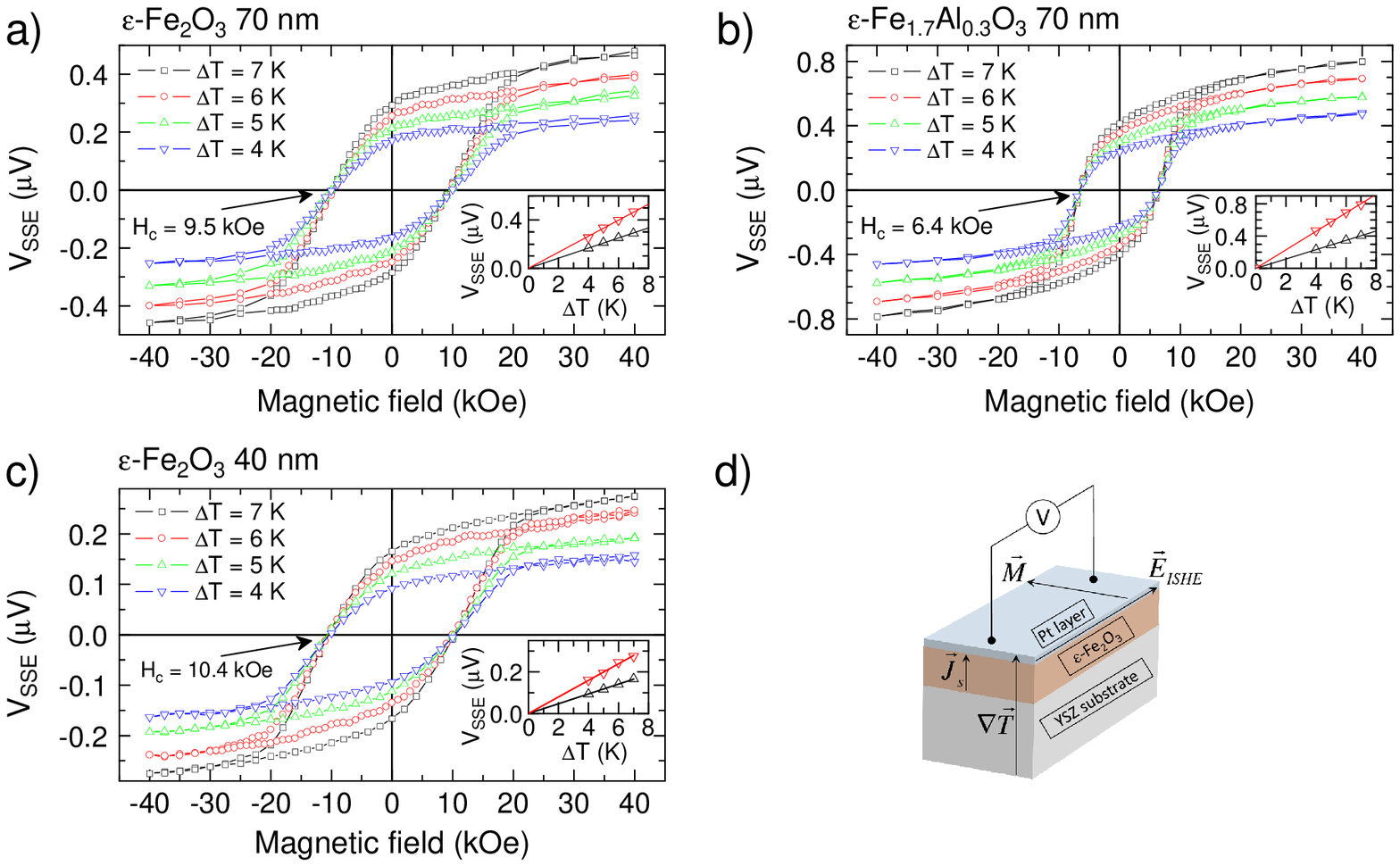} 
\caption{SSE for various temperature gradients of a) \eps-70~nm, b) \epsal-70~nm and c) \eps-40~nm. Insets show the dependence of SSE at 0~kOe ($\triangle$) and 40~kOe (\textcolor{red}{$\nabla$}).
d) Schema of the longitudinal experimental configurations. Directions of spin current ($\vec{J}_s$), temperature gradient ($\nabla\vec{T}$), magnetization ($\vec{M}$), and electrical field resulted from inverse spin Hall effect ($\vec{E}_{ISHE}$), are shown.
} \label{Fig_SSE}
\end{figure*}

The magnetic properties of the \epsx\ thin layers were characterized by magnetization loops measured at room temperature in parallel orientation with external magnetic field. In order to obtain solely the film signal, the bare YSZ substrate was measured separately and the magnetization loops have been corrected for the diamagnetic contribution of the substrate.

The hysteresis loop for \eps-70~nm is displayed in Fig.~\ref{Fig_Mag}a.
The shape of the loop is typical for ferrimagnetic materials with a pronounced high-field induced paraprocess above 20~kOe.
No distortion of the loop, which would indicates a presence of magnetic impurities, is observed.
The coercive field is $H_c=11.6$~kOe.
Although this value is smaller than $H_c=20$~kOe reported for \eps\ in the form of nanoparticles, \cite{RefJin2004ADVMAT16_48,RefKubickova2016HFI237_159}, it is larger than values reported so far for thin films, where maximum 0.8~kOe was found \cite{RefCorbellini2017SCRMAT140_63,RefGich2010APL96_112508}.
The magnetic remanence 32~\emucm\ is comparable to the value reported for nanoparticles (around 35-40~\emucm) \cite{RefJin2004ADVMAT16_48,RefKubickova2016HFI237_159}

In distinction to \eps, the hysteresis loop of \epsal\ thin layer exhibits a slightly constricted shape, see the Fig.~\ref{Fig_Mag}b.
Constricted shape can be explained by a coexistence of two ferro/ferri-magnetic phases (soft and hard), in this case by a presence of a secondary minor phase with soft magnet properties. Among iron oxides, the possible soft ferrimagnets are \feomagh\ and \feomagn, considered also as magnetic impurities in \cite{RefGich2010APL96_112508,RefCorbellini2017SCRMAT140_63}. Both these phases have spinel structure and have similar magnetic properties, with saturated magnetization $\sim$400 and $\sim$300~\emucm\ for \feomagh\ and \feomagn, respectively, see \textit{e.g.} the hysteresis loops of thin layers of \feomagh\ and \feomagn\ in \cite{RefJimenez-Cavero2017APLMAT5_26103}.
Since the saturated magnetization of both spinel phases is several times higher than magnetization of \eps, the correction of 1~volume\% of \feomagh\ or 1.3~volume\% of \feomagn\ was sufficient to remove the inflections in the hysteresis loop. This low content of secondary phase was not detected by XRD. After the correction, the coercive field of \epsal\ is $H_c=6.5$~kOe, which is smaller compared to \eps\ thin layer, as expected for Al-substituted phases \cite{RefCorbellini2017SCRMAT140_63}.
The hysteresis loop for \eps-40~nm is displayed in the Fig.~\ref{Fig_Mag}c.
It exhibits a more constricted shape with apparently $H_c=8.2$~kOe.
After the correction of 1.8~volume\% of \feomagh\ or 2.3~volume\% of \feomagn, the coercive field is $H_c=11.0$~kOe, comparable with 70~nm film.


\begin{figure}
\centering
\includegraphics[width=1.00\columnwidth,viewport=0  100 535 780,clip]{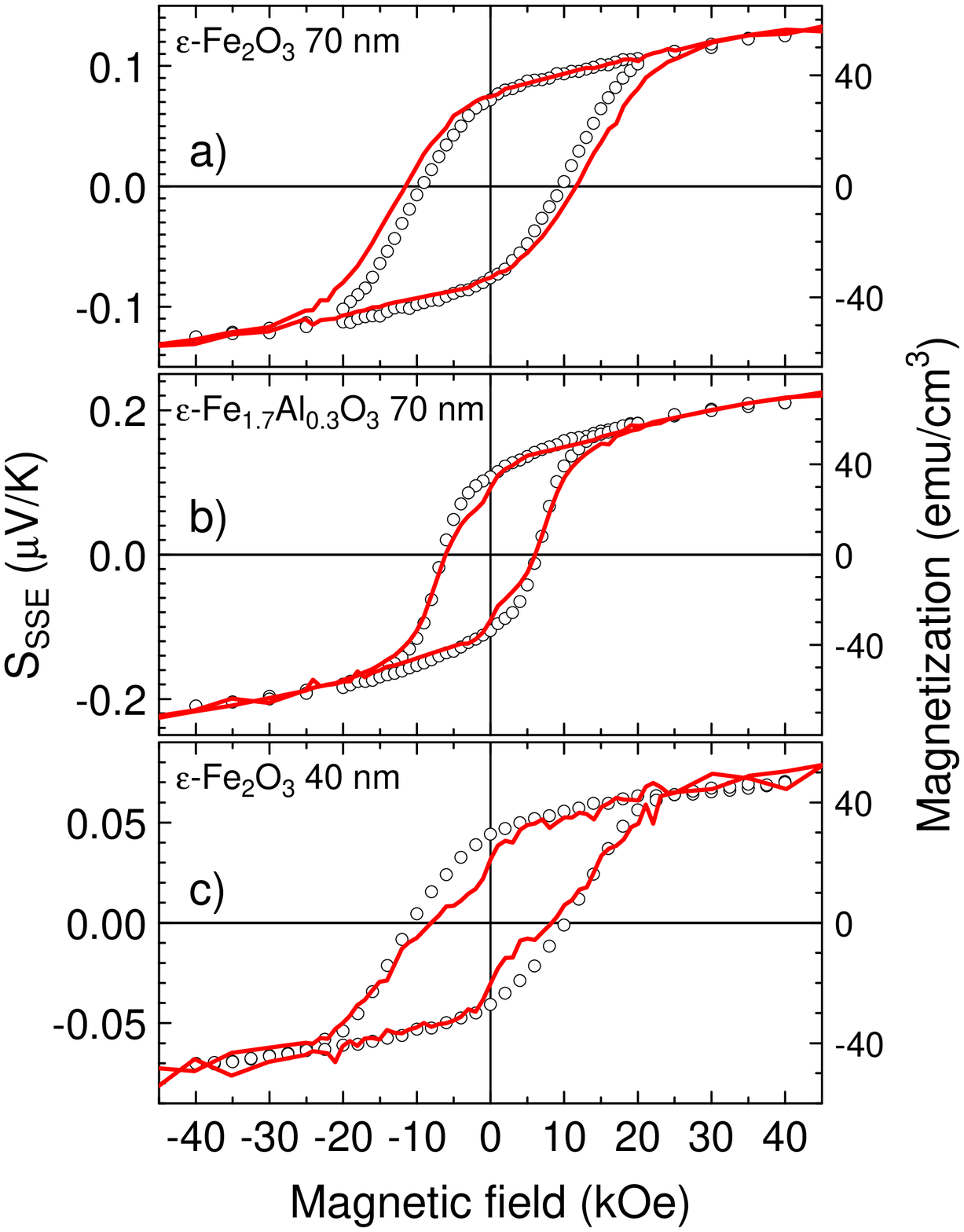} 
\caption{Comparison of  magnetization and SSE hysteresis of \eps-70~nm, \epsal-70~nm and \eps-40~nm.} \label{Fig_Mag_SSE}
\end{figure}

The spin Seebeck signal was measured in the longitudinal configuration depicted in Fig.~\ref{Fig_SSE}d, in which the directions of the magnetic field is parallel with the film surface, the same as in magnetic measurements.
The spin Seebeck signal of \epsx\ for various temperature differences $\triangle T$ at room temperature is displayed in Figs.~\ref{Fig_SSE}a-c. The shape of the spin Seebeck signal in dependence on the magnitude and polarity of the magnetic field basically resemble the magnetization loops. The spin Seebeck loops exhibit similarly high coercive fields and continuous increase of the signal at high field.
Both the spin Seebeck signal determined at zero field, \textit{i.e.} remanent spin Seebeck signal, and determined at maximum field 4~T, linearly increase with temperature difference $\triangle T$, see the insets of the Fig.~\ref{Fig_SSE}a-c.
The coercive field is independent on $\triangle T$.


The similarity of the magnetic and SSE hysteresis loops is demonstrated in the Fig.~\ref{Fig_SSE}.
The shape of the hysteresis loops are practically the same for \eps-70~nm layer, see the Fig.~\ref{Fig_SSE}a. The coercive field is 9.5~kOe for, which slightly lower than that of magnetization loops.
In the case of \epsal-70~nm layer the coercive field $H_c$ is to 6.4~kOe, in good agreement with magnetization $H_c$ after correction, see the Fig.~\ref{Fig_SSE}b. However, in distinction to constricted magnetization loop, the spin Seebeck loop exhibits no deformation.
This difference between magnetization and spin Seebeck loops is also observed for \eps-40~nm layer, see Fig.~\ref{Fig_SSE}c. Although the magnetic hysteresis loop is markedly constricted, there is no indication of deformation of the spin Seebeck loop by a secondary phase.

This is rather unexpected, since both \feomagn\ and \feomagh\ phases, which are considered as possible secondary magnetic phases, were proved to exhibit significant spin Seebeck signal \cite{RefRamos2013APL102_072413,RefJimenez-Cavero2017APLMAT5_26103}.
The accurate comparison of SSE measured by different laboratories is difficult, since in most of the experimental setups the temperature sensors measuring the temperature difference $\triangle T$ describes not merely the thermal characteristics of the studied layered material but the whole measurement cell instead, which makes the quantity in units of \muvk\ physically irrelevant to the SSE itself \cite{RefSola2015JAP117_17C510,RefSola2017SCIREP7_46752,RefHirschner2017PRB96_064428}. Nevertheless, as the SSE in \epsx\ is about 10$\times$ smaller than in \feomagn\ or \feomagh, so even taking into account this uncertainty, it could be expected that their presence should affect SSE signal similarly as magnetization.

A different shape of the magnetization and SSE loops was also observed for 1~mm thick (Mn,Zn)Fe$_2$O$_4$ sample in \cite{RefUchida2010APL97_262504}, where SSE exhibited hysteresis with $H_c \sim 1$~kOe in contrast to magnetization without hysteresis. It was explained by the authors, that the voltage signal reflects only the interface magnetization information, which can be different from magnetization for the bulk in general. However, in our case of 40 and 70~nm thin films, the SSE is generated from the whole film volume, that is confirmed by the proportional increase of the signal with the film thickness.

In our case the explanation is probably related to percolation of the spin current. If the minority ferrimagnetic phase is not present at the thin film surface, \textit{i.e.} near the Pt layer, the spin current of the minority phase do not percolate up to the Pt layer, therefore do not contribute to SSE. However, further studies are needed to confirm this hypothesis.

\section{Conclusions}

Spin Seebeck effect has been investigated in Pt/\epsx\ bilayers in longitudinal configuration. The \epsx\ ($x=0$ and 0.3) thin layer with $40-70$~nm thickness were deposited by a spin-coating method on Y:ZrO$_2$(100) substrates. The prepared layers are highly oriented with the easy magnetic $a$-axis parallel to the film surface. The magnetic hysteresis loops measured at room temperature with magnetic field parallel to the layer exhibit coercive fields up to 11.6~kOe, which is so far the highest value measured for thin layer samples. The shape of the spin Seebeck hysteresis loops is similar to the shape of magnetization for single phase layers with coercive field around 10~kOe. A difference in encountered in the case of layers with a small amount ($1-2$~volume\%) of secondary soft ferrimagnetic phase, which is revealed by a constricted shape of magnetization loops, in contrast to spin Seebeck loops, where no constriction is observed.
This difference is probably related to a percolation of the spin current. If the minority ferrimagnetic phase is not near the thin film surface, \textit{i.e.} near the Pt layer, the spin current of the minority phase do not percolate up to the Pt layer, therefore do not contribute to spin Seebeck signal.


\textbf{Acknowledgement}.
This work was supported by Project No.~16-04340S of the Czech Science Foundation.


\end{document}